\documentclass[conference]{IEEEtran}

\usepackage{amsmath}
\usepackage{graphicx}
\usepackage{epstopdf}

\newcommand{\beq}{\begin{equation}}
\newcommand{\eeq}{\end{equation}}

\begin{document}
%
\title{Accurate Switching Currents Measurements in Quantum Washboard Potential}

\author{
\IEEEauthorblockN{Vincenzo Pierro}
\IEEEauthorblockA{\small 
Dept. of Engineering\\
University of Sannio\\
C.so Garibaldi 107, 
82100 Benevento, IT\\
Email: pierro@unisannio.it}
\and
\IEEEauthorblockN{Giovanni Filatrella}
\IEEEauthorblockA{\small
Dept. of Science and Technology \\
 University of Sannio \\ Via Port'Arsa 11, 82100 Benevento, IT \\
 Email: filatrella@unisannio.it \\ 
}
}

\maketitle

\begin{abstract}
We tackle the problem of accurate simulations  of switching currents arising from tunnel events in the washboard potentials associated to Josephson junctions.
The measurements of the probability distribution of the switching currents is essential to determine the quantum character of the device, and therefore is at the core of technological applications, as Josephson junctions, that have been proposed for quantum computers.
In particular, we show how to accurately calibrate the parameters of the boundary conditions to avoid spurious reflections of the wavefunction from the finite border of numerical simulations. 
The proposed approximate numerical scheme exploits a quantum version of a prefect matched layers  for the boundary problems associated with this class of potentials.
Thus, we employ the analogous of a well established electromagnetic method to deal with radiation in mesoscopic quantum systems.
Numerical simulations demonstrate that the known analytic results are well recovered in the appropriated limits of quantum measurements.
We also find that a relaxation time shows up in the dynamics of the quantum evolution in between two consecutive measurements.
\end{abstract}
\IEEEpeerreviewmaketitle

\section{Introduction}
Tunnel events are related to the probability to locate the quantum particle in a region that is classically forbidden. 
This phenomenon occurs in many systems, for example Josephson Junctions (JJ) \cite{Barone82},\cite{Tinkham96}, and it is in particular relevant to understand new devices, based on novel materials \cite{Massarotti15}, as graphene based superconducting devices \cite{Lee11,Coskun12}. 
Tunnel in JJ occurs in the form of {\em switching currents}, or the sudden appearance of a voltage at a bias current value where the junction is expected to be in the superconducting zero voltage state. 
This passage from the superconducting to the resistive state corresponds, in the mechanical interpretation of the governing potential, to the passage from the confined to the {\em running solution}, despite the presence of an energy barrier higher than the available energy  \cite{Tinkham96}.

The detection of tunnel events, as opposed to thermal activation \cite{Guarcello15}, is of potential interest in quantum measurements, for example in the measurements connected with quantum computing \cite{Makhlin01,Price10}.
Also the development of advanced gravitational wave detectors \cite{Abbott16}, working beyond the thermal noise limit \cite{Agresti06,Villar10} and near the quantum regime of test masses \cite{Addesso13}, is connected to the detection of tunnel events.
To reproduce the quantum behavior in numerical simulations requires to deal with boundary conditions, in order to return accurate tunnel rates.
In fact a problem arises in treating tunnel for potentials unbounded from below, as the washboard or the cubic \cite{Caliceti80,Alvarez88}.
For these potentials, when the system escapes from the trapped state, it moves to a {\it running solution}, that physically corresponds to the absorption of the particle. 
It is interesting to mention the electromagnetic analog of tunnel phenomenon in unbounded from below potentials.
Localized states correspond to a radiation trapped in between two purely dielectric surfaces, tunnel to the passage of radiation through transparent barriers.
The absorption of a particle outside the barrier corresponds to an electromagnetic solution that radiates towards infinity, while  the energy is constantly absorbed by the boundary \cite{Taflove95}.
The purpose of this paper is to illustrate how to exploit this analogy to accurately calculate the rate of tunnel events for time dependent potentials, inasmuch analytical solutions are not available.
For sake of illustration, we will focus on Josephson junctions, even if the method is general and could in principle be employed for a generic cubic type potential.
The work is organized as follows. 
In Section \ref{sect:model} we describe the mathematical model of the quantum system. 
In Sect. \ref{sect:measure} we propose a technique to deal with quantum measurements and to retrieve accurate tunnel probabilities as a function of the bias current. 
Sect. \ref{sect:conclusions} concludes.

\section{Mathematical model}
\label{sect:model}
Tunnel occurs when a system representative point overcomes an energy barrier that is higher than the available energy.
If the process is to be resolved in time, it amounts to the detection of the instant when a particle has escaped the trapping potential and has arrived at a point beyond the barrier. 
Arrival times in quantum measurements  pose some physical difficulties \cite{Anastopoulos06}, especially  for potential unbounded from below \cite{Andersen13}.
In particular for the tilted washboard potential of JJ, the passage to an accelerated  running state is problematic, in that the associated state is not an eigenfunction of the Hamiltonian \cite{Paraoanu05}.
Consequently, it cannot be translated in the standard von Neumann ideal model of quantum measurements. 
The problem is further complicated if the potential is time dependent.
In this case, of course, one cannot use stationary solutions, and the problem is mostly treated with numerical simulations.
Numerical simulations are, however, difficult, for one should ensure that the absorption at a finite boundary (in practice: at the limit of the numerical domain of integration) mimics the absorption at infinity of the corresponding mathematical model. 
The first part of this paper is devoted to such a problem, prior to the investigation of quantum dynamics in between two measurements.

\begin{figure}[t]
\centering
\includegraphics[width=2.5in]{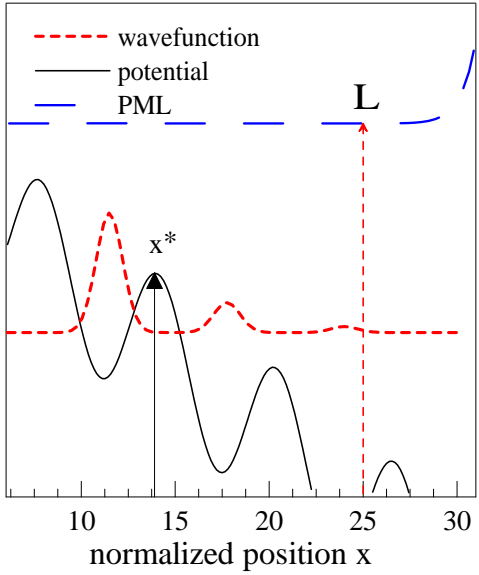}
\caption{Sketch of tunnel events. The solid line denotes the washboard potential, while the blue long dashed line is the imaginary part of the absorber, corresponding to an electromagnetic  Perfect Matched Layer. The red dashed curve is  the square modulus of the wavefunction. The coordinate of the maximum of the trapping potential well is denoted by $x^\star$, while $L$ denotes the numerical integration domain.
}
\label{fig:potential}
\end{figure}

To schematically illustrate the measurement procedure that we want to reproduce, we report  in Fig. \ref{fig:potential} a quantum particle in a washboard, for example a Josephson junction or any other system described by a (tilted) periodic potential. 
The corresponding dynamics is given by the solution of the Schr\"odinger equation \cite{Tinkham96}:
\beq
i\hbar \frac{\partial \psi}{\partial t'} = \left[ - \frac{\hbar ^2}{2M}\frac{\partial^2}{\partial x^2} -  E_J \left(  \cos (x ) +  \gamma(t')x \right)  \right] \psi,
\label{eq:JJSE}
\eeq
where $M$ is the particle mass and $E_J$ is some characteristic energy. 
In normalized units \cite{Andersen13} it reads:
\begin{eqnarray}
i \frac{\partial \psi}{\partial t} &=& \left[ -\frac{1}{2}\frac{\partial^2}{\partial x ^2} -  V_0 \left( \cos\left(x \right) +  \gamma (t) x \right) \right] \psi  = \nonumber \\
&=& \left [ -\frac{1}{2}\frac{\partial^2}{\partial x ^2} + U_0(x)  \right] \psi ,
\label{eq:JJSEn}
\end{eqnarray}
\noindent In the Josephson junction realization of the system, time is normalized to $\hbar /C\left ( \Phi_0/2\pi \right)^2$, where $C$  is the junction  capacitance, and $\Phi_0= h/2e$ is the flux quantum.
The term $\gamma(t)= I(t)/I_0$ is the normalized (respect to  $I_0$, the critical current of the JJ) applied current $I (t)$; this current is ramped in a time $T$ from $\gamma(0)=0$ to $\gamma(T)=1$. 
The parameter $V_0=E_J/E_C$ is the normalized maximum energy barrier that results from the ratio between the Josephson energy $E_J=I_0\Phi_0/2\pi$ and the Coulomb energy $E_C=\hbar^2/C\left(\Phi_0/2\pi\right)^2$ . 
Equation (\ref{eq:JJSEn}) stems from the washboard potential:
\beq
U_w (x)= V_0   \left( \cos x + \gamma x  \right).
   \label{eq:potential_washboard}
\eeq
\noindent 
for the standard Josephson junctions, or to some more complicated potentials for graphene based junctions \cite{Valenti14}.
The potential (\ref{eq:potential_washboard}) can be (locally) approximated by the cubic  potential \cite{Caliceti80}:
\beq
U_c (x) = - g  x^3 + \frac{1}{2} \omega_p^2 x^2 
   \label{eq:potential_cubic}
\eeq
The physical interpretation of the coefficients $g$ and $\omega_p$ in Eq.(\ref{eq:potential_cubic}) is that the minimum occurs at $x=0$ and the energy barrier depends upon the parameters $ \Delta U =\omega_p^6/(54g^2)$. 
Thus, one can calibrate the coefficients in Eq.(\ref{eq:potential_cubic}) to reproduce the main physical properties of the washboard potential sketched in Fig. \ref{fig:potential} for $V_0 \gg 1$. 

To increase the normalized bias current $\gamma$ amounts to tilt the potential associated to Eq.(\ref{eq:potential_washboard}), while the corresponding barrier (normalized respect to $E_C$):
\beq
\label{eq:potential}
\Delta U =2 V_0 \left[ \sqrt{1-\gamma}-\gamma  \arccos{\gamma}\right]
\eeq
decreases and eventually, close to $\gamma=1$, becomes small enough to make the tunnel probability  sizable. 
To detect tunnel events in washboard potentials corresponds to the possibility to observe the passage of the system across the confining energy barrier, i.e. from the trapping region to the outside region  \cite{Valenti14,Wallraff03a,Wallraff03b,Augello09}.

The effect is evident in Fig. \ref{fig:potential} as a small bump of the quantum wavefunction in the subsequent minima of the potential.
The evolution of the wavefunction $\psi$ is analogous to the electromagnetic propagation of radiation in an inhomogeneous medium with open boundary conditions.
Tunnel events and passage to a running solution is interpreted in the electromagnetic analogy as radiation towards infinity.
Runaway resonant  solutions \cite{Garcia97,Cordero11}, in the frequency domain, can be modeled by an absorbing boundary condition of the type 
\beq
\lim_{x \rightarrow \infty} \left( \frac{\partial  \hat{\psi}(x,\omega)}{\partial x} -ik(\omega)\hat{ \psi}(x,\omega) \right)  = 0,
\label{eq:bc_absorbing}
\eeq
with an appropriated choice of $k$ ($\hat{\psi}$ is the Fourier transform of the wavefunction $\psi$). 
This boundary condition has the advantage that can be mathematically handled, and for instance is employed to retrieve the WKB approximation \cite{Tinkham96}. 
However, as mentioned, it only models the stationary solutions; as such, when the bias current $\gamma(t)$ of Eq.(\ref{eq:JJSEn}) depends upon the time, it is not fully appropriated. 
Ideally, for a time dependent problem one could just impose the conditions in a point infinitely faraway, and set:
\beq
\psi  \left(t,  x = \pm \infty \right) =  0.
\label{eq:bc_zero}
\eeq
Zero boundary conditions are valid for any time $t$, but the time dependent Eq.(\ref{eq:JJSEn}) cannot be solved with these conditions. 
It is therefore necessary to retrieve the solution by means of numerical simulations, that of course cannot handle infinite distance. 
The practical implementation is to impose that the solution is absorbed at a finite distance, say $L$.
The technical problem with a zero boundary at a finite distance $L$ is the following: a wave that encounters a zero condition, is reflected by the discontinuity (the analogous of a wave reflected by an abrupt change of the reflectivity index). 
This reflection is unphysical, only due to the finite condition at $L$ and leads to inaccuracies in the tunnel rate.
The problem is particularly deceptive for accurate measurements of quantum events, as will be discussed in the following Section.  
To avoid such spurious reflection, one can use an absorbing imaginary potential that cancels the reflected waves.
Following the electromagnetic analogy, we propose to employ boundary conditions of the type:
\beq
U_{abs}= i \exp \left[ \frac{  x-x_0 }{ l_{ext}} \right] \left[ A \left( x-x_0\right) \right]^6.
\label{eq:absorbing}
\eeq
where $x_0$ is the absorbing potential minimum, $A$ the amplitude and $l_{ext}$ some appropriated length over which the imaginary part decays, see Fig. \ref{fig:potential}.
This imaginary part corresponds, in the electromagnetic analogy, to a Perfect Matched Layer (PML) that, ideally, absorbs the incoming waves without significant reflection \cite{Taflove95,Xu07}. 
The expression (\ref{eq:absorbing}) ensures that in the matching point $x_0$ the potential tends to zero, together with the first six derivatives, in accordance with the standard PML layers design rules.
The length $l_{ext}$ and the amplitude $A$ should be chosen to absorb the quantum particles within the finite domain of numerical integration.
Absorption occurs when the energy of the incoming particle is matched by the appropriated amplitude of the imaginary potential. 
Thus, the smooth behavior guarantees that each component is blocked in a different point of the potential, and backscattering is minimized.
To do so, in Fig. \ref{fig:transient} we have employed $A=10^{-3}$, $l_{ext} = 10^3$, $L-x_0 = 10^3$.
With these parameters we have found that the results are in agreement with the WKB approximation (in the stationary limit) and the backscattering is negligible.
It is important to underline that the imaginary part of the potential entails a dissipation that emulates on a finite domain the effect of open boundary radiative conditions at infinity. 
Thus, it is not unphysical that the total norm of the wavefunction is not any more preserved in the integration domain.
\begin{figure}[t]
\centerline{\includegraphics[scale=0.45]{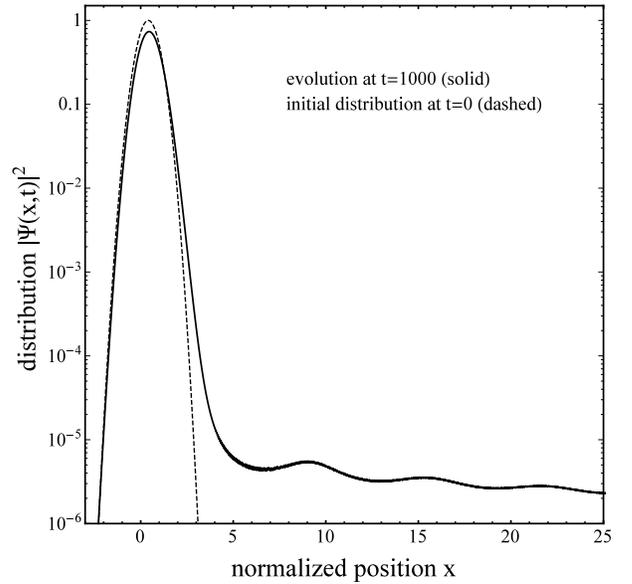}}
\caption{
Time evolution of the wavefunction according to Eq.(\ref{eq:JJSEn}).
The thin dashed line denotes the initial condition for $t=0$, while the thick line denotes the solution at $t=1000$. The bias current is $\gamma =0.4$, and the boundary condition are given by PML as per Eq.(\ref{eq:absorbing}), with parameters:  $A=10^{-3}$, $l_{ext} = 10^3$, $L-x_0 = 10^3$.
}
\label{fig:distribution}
\end{figure}

We introduce the probability to locate the system in any finite domain $P(x<\infty)$:
\beq
P(x<\infty)  = \lim_{L \rightarrow \infty}  \int_{-L}^{+L} |\psi(x)|^2 dx,
\label{eq:P}
\eeq
where the prescription to evaluate the improper integral is that at $L$ there is a perfect absorber that cancels the reflected waves (i.e., the radiation terms).
Physically, $P(x<\infty)$ corresponds to the portion of the probability that is not radiated in the space outside the domain of integration.
Put it another way, the waves are in fact traveling in the region $x>L$, and therefore in the numerical domain of integration $-L <  x < +L$ there is a probability to find the particle that is less than $1$.

\section{Measuring tunnel events}
\label{sect:measure}

\begin{figure}
\centerline{\includegraphics[scale=0.45]{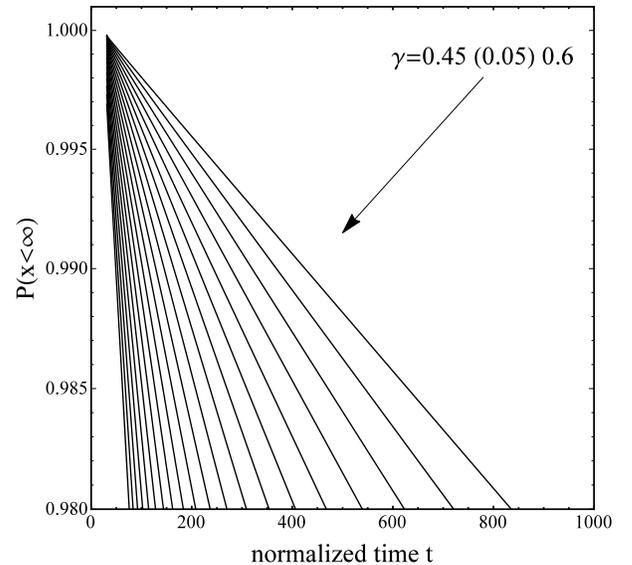}}
\caption{
The probability to locate the system in the integration domain as a function of the time. 
The data demonstrate, for different values of the bias current, that the solution norm is not conserved in the integration domain $x < \infty$.
The decaying is, in the long run, exponential as expected in the standard WKB approximation. 
The bias current is increased from the value $\gamma=0.45$ to $\gamma = 0.6$ by steps of amplitude $\Delta \gamma = 0.05$.
The PML parameters are the same as in Fig. \ref{fig:distribution}.
}
\label{fig:transient}
\end{figure}

As discussed in the previous Sections, tunnel events occur when, during the time evolution of  the  Schr\"odinger   Eq. (\ref{eq:JJSEn}), one encounters a finite probability to find a particle outside the potential well \cite{Dhar15,Matta15}. 
An example is shown in Fig. \ref{fig:distribution}, that displays the evolution of an initially localized solution, from $t=0$ to the time $t=1000$. 
From the Figure it is apparent that the evolved solution exhibits a nonzero probability flux, corresponding to the radiation described by Eq.(\ref{eq:bc_absorbing}) . 
From Fig. \ref{fig:distribution} it is also evident that the tilted potential (see Fig. \ref{fig:potential}) ensures that the probability current flows towards $x=\infty$, while the contribution of the flux towards $x = -\infty$ is negligible. 
This probability current is therefore influenced by the boundary conditions, and crucially depends, in simulations, by the possibility to correctly approximate zero boundary conditions (\ref{eq:bc_zero}) in a finite domain with the absorbing PML, as discussed in Sect. \ref{sect:model}.
To check the validity of this approximation we have focused on the probability flux, that is connected with the abovementioned probability $P(x<\infty)$ given by Eq.(\ref{eq:P}), in Fig. \ref{fig:transient}.
In this Figure we display $P(x<\infty)$, the probability to locate the system in the integration domain $x < \infty$ as a function of the time for different values of the bias current.
The behavior is compatible with a decaying resonance due to the absorbing boundary condition at infinity , as per Eq.(\ref{eq:bc_absorbing}). 
The rate of absorption also depends upon the bias current $\gamma$, and becomes higher for higher bias values, see Fig. \ref{fig:transient}.
The decrease of the norm in the integration domain is not surprising, inasmuch for finite bias current the stable trapped solution (the thin line of Fig. \ref{fig:distribution}) is not a bona-fide stable eigenfunction.
The data confirm that the decaying rate is exponential, and in reasonable agreement with the WKB approximation. 
It is therefore possible to control the PML, Eq.(\ref{eq:absorbing}), absorber to guarantee that the boundary condition (\ref{eq:bc_absorbing}) correctly describes the {\em probability radiation} towards infinity.

\begin{figure}
\centerline{\includegraphics[scale=0.45]{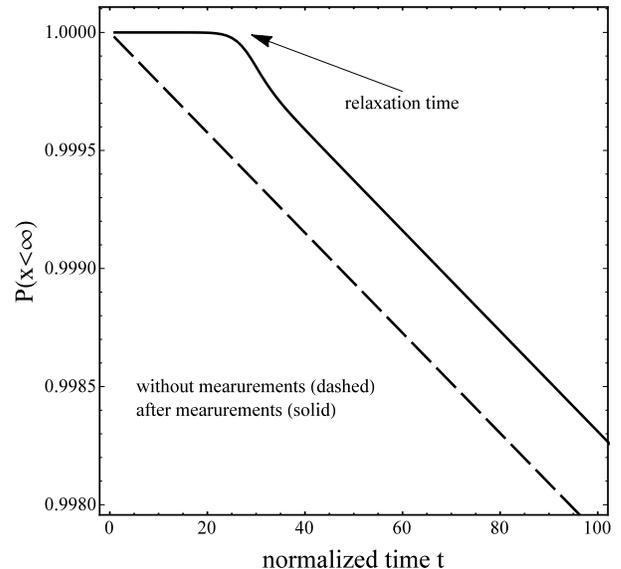}}
\caption{
The effect of the absorbing boundary conditions. 
The figure shows the behavior of the probability to locate the system in the integration domain. 
The dashed line corresponds to the case when the initial condition is the lower  resonance of the potential, while the solid curve indicates the same quantity after a quantum measurement has occurred.
We note that in the first stage of the evolution from the measured state the absorber does not change this probability and the spatial distribution of the solution stays close to the initial distribution.
When the relaxation time has passed, the probability distribution approaches the lower resonant solution characterized by a decaying rate. 
The decaying part of the solution has the same slope that is compatible with the WKB approximation.
The normalized bias current reads $\gamma=0.4$. 
The PML parameters are the same as in Fig. \ref{fig:distribution}.
}
\label{fig:absorber}
\end{figure}

The behavior of $P(x<\infty)$ between two measurements, for a constant value of the bias current $\gamma$, is displayed in Fig. \ref{fig:absorber}. 
The Figure demonstrates that the solution does not reach the asymptotic rate immediately, inasmuch for a while it persists in the original state. 
In Fig. \ref{fig:absorber} we compare the asymptotic decaying solution (dashed line) with the evolution  from the initial state after the measurement (solid line).
The displayed time evolution of {\em radiated probability} exhibits a knee, i.e. an initial flat part (a relaxation time) where this probability flux is still close to zero.
Such relaxation time is the time needed for a solution, after a measurement where tunnel has not occurred, to reach the asymptotic state.
Being a consequence of a measurement, it is a signature of the quantum behavior.
In fact, in classical measurements there is no consequence of a null measurement (the solution is found localized, and has {\it not} tunneled).
Instead, in quantum measurements a null measurement has the effect to impose the corresponding solution, and it will take some time (about $20$ normalized units in Fig. \ref{fig:absorber}) to reach again the asymptotic resonance.

\section{Conclusion}
\label{sect:conclusions}
To deal with open boundaries of a mesoscopic quantum system, modelled by a washboard potential, we have transposed a well known technique used in electromagnetic applications, the PML\cite{Taflove95}.
We have in fact considered a Josephson junction that is progressively biased until the representative coordinate tunnels and a switch current is measured.
To simulate tunnel events, we have treated the boundary conditions using the analogy with electromagnetic radiation with the choice of an appropriated quantum version of a PML, i.e. an imaginary absorbing potential, Eq.(\ref{eq:absorbing}).
We have observed that a PML can be employed also in accurate numerical simulations of time dependent potentials that are not analytical treatable.
The proposed numerical scheme allows accurate quantum simulations of the transient dynamics between two measurements.
The emergence of a relaxation time points towards possible deviations from the standard WKB result \cite{Tinkham96} for switching current distributions \cite{unpublished}.


\section*{Acknowledgment}
We appreciate valuable discussions with I.M. Pinto and financial support from 
 Programma regionale per lo sviluppo innovativo delle filiere Manifatturiere strategiche della Campania— Filiera WISCH Progetto2: Ricerca di tecnologie innovative digitali per lo sviluppo sistemistico di computer, circuiti elettronici e piattaforme inerziali ad elevate prestazioni ad uso avionico.

V.P. acknowledges partial financial support from I.N.F.N. Sez. Napoli \& Salerno. 
G.F. acknowledges partial financial support from PON Ricerca e Competitivit\`{a} 2007-2013
under grant agreement PON NAFASSY, PONa3\_00007


\begin{thebibliography}{99}

\bibitem{Barone82} A. Barone and G. Patern\`o, "Physics and application of the Josephson effect", New York, Wiley, 1982.
\bibitem{Tinkham96} M. Tinkham, "Introduction to Superconductivity: second edition," New York, Dover, 1996.
\bibitem{Massarotti15} D. Massarotti, A. Pal, G. Rotoli, L. Longobardi, M.G. Blamire, and F. Tafuri,
"Macroscopic quantum tunnelling in spin filter ferromagnetic Josephson junctions,"
{\it Nature Comm.}, Vol.  { 6}, p. 7376, 2015.
\bibitem{Lee11}G. Lee, D. Jeong, J. Choi, Y. Doh, and H.  Lee, "Electrically Tunable Macroscopic Quantum Tunneling in a Graphene-Based Josephson Junction," {\it Phys. Rev. Lett.}, Vol. 107, p. 146605, 2011.
\bibitem{Coskun12}U. C. Coskun, M. Brenner, T. Hymel, V. Vakaryuk, A. Levchenko, and A. Bezryadin, "Distribution of Supercurrent Switching in Graphene under the Proximity Effect," {\it Phys.  Rev. Lett.}, Vol. 108, p. 097003, 2012.
\bibitem{Guarcello15} C. Guarcello, D. Valenti, and B. Spagnolo, "Phase dynamics in graphene-based Josephson junctions in the presence of thermal and correlated fluctuations,"
{\it Phys. Rev. B}, Vol.  {92}, p. 174519, 2015.
\bibitem{Makhlin01}Y. Makhlin, G. Sch\"n, and A. Shnirman, "Quantum-state engineering with Josephson-junction devices," {\it Rev. Mod. Phys.}, Vol.  73, p. 357, 2001,
\bibitem{Price10}
A. N. Price, A. Kemp, D.R. Gulevich, F.V. Kusmartsev,  and A.V. Ustinov, "Vortex qubit based on an annular Josephson junction containing a microshort," {\it Phys. Rev. B}, Vol. { 81}, p. {014506}, 2010.
\bibitem{Abbott16}V. Pierro, LIGO Scientific Collaboration and Virgo Collaboration, 
"Observation of Gravitational Waves from a Binary Black Hole Merger,"
{\it Phys. Rev. Lett.}, Vol. 116, p. 061102, 2016.
\bibitem{Agresti06} J. Agresti, G. Castaldi, R. DeSalvo, V. Galdi, V. Pierro,
and I. M. Pinto, "Optimized multilayer dielectric mirror
coatings for gravitational wave interferometers," {\it Proc.
SPIE}, Vol.  6286, p. 628608, 2006.
\bibitem{Villar10} A. E. Villar, E. D. Black, R. DeSalvo, K. G. Libbrecht, C. M. N. Morgado, L. Pinard, I. M. Pinto, V. Pierro, V. Galdi, M. Principe, and I. Taurasi., "Measurement of thermal noise in multilayer coatings with optimized layer thickness," {\it Phys. Rev. D}, Vol.  81, p. 122001, 2010.
\bibitem{Addesso13}P. Addesso, V. Pierro, and G. Filatrella, "Escape time characterization of pendular Fabry-Perot," {\it European Physics Letter}, {Vol. 101},  p. 200051, 2013. 
\bibitem{Caliceti80} E. Caliceti, S. Graffi, and M. Maioli, "Perturbation  Theory of Odd  Anharmonic Oscillators,"  {\it Commun. Math. Phys.}, Vol.  {75}, p. 51, 1980.
\bibitem{Alvarez88} G. Alvarez, "Coupling-constant behavior of the resonances of the cubic anharmonic oscillator," {\it Phys. Rev. A}, Vol. { 37}, p. 4079, 1988.
\newpage
\bibitem{Taflove95} A. Taflove, "Computational electrodynamics, the finite difference time-domain method," Boston, Artech House, 1995.
\bibitem{Anastopoulos06} C. Anastopoulos and N. Savvidou, "Time-of-arrival probabilities and quantum measurements," {\it J. Math. Phys}, Vol.  { 47}, p. 122106, 2006; "Time-of-arrival probabilities and quantum measurements. II. Application to tunneling times," {\it ibid. }, Vol.   { 49}, p. 022101, 2008.
\bibitem{Andersen13}C.K. Andersen and K. M\o{}lmer, " Effective description of tunneling in a time-dependent potential with applications to voltage switching in Josephson junctions," {\it Phys. Rev. A}, Vol. { 87}, p. 052119, 2013.
\bibitem{Paraoanu05}G. S. Paraoanu, "Running-phase state in a Josephson washboard potential," {\it Phys. Rev. B}, Vol. { 72}, p. 134528, 2005.
\bibitem{Valenti14} D. Valenti, C. Guarcello, and B. Spagnolo, ``Switching times in long-overlap Josephson junctions subject to thermal fluctuations and non-Gaussian noise sources'', {\it Phys. Rev. B}, Vol.  89, p. 214510, 2014.
\bibitem{Wallraff03a}
A. Wallraff, A. Lukashenko, C. Coqui, A. Kemp, T. Duty, and A. V. Ustinov,
"Switching current measurements of large area Josephson tunnel junctions,"
{\it Rev. Sci. Instr.}, Vol. 74, p. 3740, 2003.
\bibitem{Wallraff03b} A. Wallraff, A. Lukashenko, J. Lisenfeld, A. Kemp, M. V. Fistul, Y. Koval, and A. V. Ustinov, "Quantum dynamics of a single vortex," Nature Vol. 425, p. 155, 2003.
\bibitem{Augello09}G. Augello, D. Valenti, A.L. Pankratov, and B. Spagnolo, "Lifetime of the superconductive state in short and long Josephson junctions," {\it Eur. Phys. B}, Vol.  { 70}, p. 145, 2009.
\bibitem{Garcia97}G. Garc\'{i}a-Calder\'{o}n and A. Rubio, 
"Transient effects and delay time in the dynamics of resonant tunneling," {\it Phys. Rev. A}, Vol. {55}, p. 3361, 1997.
\bibitem{Cordero11} S. Cordero, G. Garc\'{i}a-Calder\'{o}n, R. Romo, and J. Villavicencio, "Unified analytical description of the time evolution of decay for initial states formed by wave-packet scattering and by initial decaying states in quantum systems," {\it Phys.  Rev.  A}, Vol. 84, p. 042118, 2011.
\bibitem{Xu07} Z. Xu, H. Han, X. Wu, 
"Adaptive absorbing boundary conditions for Schr\"odinger-type equations: Application to nonlinear and multi-dimensional problems,"
{\it J. Comp. Phys.}, Vol.  225, p. 1577, 2007.
\bibitem{Dhar15} S. Dhar, S. Dasgupta, and A. Dhar, "Quantum time of arrival distribution in a simple lattice model," 
{\it J. Phys. A: Math. Theor.}, Vol. {48},  p. 115304, 2015.
\bibitem{Matta15} V. Matta and V. Pierro, 
"Sequential nonideal measurements of quantum oscillators: Statistical characterization with and without environmental coupling,"
{\it Phys. Rev. A}, Vol.  {92}, p. 052105, 2015.
\bibitem{unpublished} V. Pierro and G. Filatrella, unpublished.




\end{thebibliography}
\end{document}